\input{aipcheck}

\def\ccpip {$CC1\pi^+$\ }
\def\ccqe {$CCQE$\ }
\def\ld{ {(\lambda)}}

\documentclass[
  ]
  {aipproc}
\layoutstyle{8x11single}

\begin{document}

\title[$Q^2$ discrepancy in the MiniBooNE \ccpip sample]{Four Momentum Transfer Discrepancy in the Charged Current $\pi^+$ Production in the MiniBooNE: Data vs. Theory}


\author{Jaroslaw A. Nowak, for the MiniBooNE Collaboration}{
  address={Department of Physics and Astronomy \\
  Louisiana State University\\ Baton Rouge, LA 70803}
}

\begin{abstract}The MiniBooNE experiment has collected what is currently the world's  largest sample of $\nu_\mu$ charged current single charged pion ($CC1\pi^+$)   interactions, roughly 46,000 events. The purity of the $CC1\pi^+$ sample is 87\% making this the purest event sample observed in the MiniBooNE detector \cite{AguilarArevalo:2008qa}. The average energy of neutrinos producing $CC1\pi^+$ interactions  in MiniBooNE is about $1~ GeV$\cite{AguilarArevalo:2008yp}, therefore the study of these events can provide insight into both resonant and coherent pion production processes. In    this talk, we will discuss the long-standing discrepancy in four-momentum  transfer observed between $CC1\pi^+$ data and existing predictions. Several  attempts to address this problem will be presented. Specifically, the  Rein-Sehgal \cite{RS81, Rein:1987cb} model has been extended to include muon mass terms for  both resonant \cite{kln, brs} and  coherent \cite{Rein2007} production. Using calculations from \cite{GS_ff}, an updated form for the vector form factor \cite {Lalakulich:2006sw} has also been  adopted.  The results of this improved description of $CC1\pi^+$ production   will be compared to the high statistics MiniBooNE $CC1\pi^+$ data and several existing parametrisations of the axial vector form factor \cite{GS_ff, Lalakulich:2006sw,Hernandez:2007qq} .
\end{abstract}

\maketitle


\section{Introduction}

In recent experiments such as K2K \cite{k2k_coh} , MiniBooNE \cite{MB::CCQE} and SciBooNE \cite{Hiraide:2008eu}  it has been observed that the description of the four-momentum transfer  in neutrino induced  interactions is not sufficient. Currently used Monte Carlo generators underestimate the neutrino induced  charged current pion production. The predicted shape is different from the observed one, especially for the low $Q^2$.   There has been an effort in the theoretical community to address  this discrepancy but mainly focused on data from bubble chamber experiments \cite{anl,bnl}.  We will present predictions within new models as applied to MiniBooNE Monte Carlo generator.

The K2K collaboration revisited the charged-current coherent pion production, concluding that  it is overestimated by existing models  and  they set an upper limit on the coherent contribution to inclusive charged-current interactions  to $0.6 \%$ \cite{k2k_coh} . The SciBooNE collaboration \cite{AguilarArevalo:2006se}  set  90\% CL upper limits for the same  ratio of $0.67\%$ and $1.36\%$  for two samples with mean neutrino energy $1.1~GeV$ and $2.2~GeV$, respectively\cite{Hiraide:2008eu}.    The MiniBooNE collaboration has observed a discrepancy between Monte Carlo prediction and data for the four-momentum transfer in the results for \ccqe interaction on carbon.   In order to fix this problem the MiniBooNE collaboration set  an effective value of axial-vector mass at  $M_A^{eff} =1.23\pm 0.20 GeV$, and introduced an effective parameter $\kappa=1.019 \pm 0.011$, which modifies the Pauli-suppression and reduces the discrepancy between the data and Monte Carlo prediction \cite{MB::CCQE}.

The prediction for  a neutrino interactions in the MiniBooNE experiment is based on the  Monte Carlo generator Nuance v3 customized for  a carbon nucleus \cite{Casper:2002sd} .  In the Nuance generator the resonance \ccpip production is describe by the Rein-Sehgal model \cite{RS81}, which is  based on the FKR model \cite{Feynman:1971wr,ravndal72,ravndal73}  and describes pion production by excitation of 18 resonances with a cut at $W=2~GeV$ in hadronic invariant mass. The resonance production is described in terms of helicity  amplitudes and the cross section contains  interference terms and nonresonant background. The vector and axial-vector form factors have a similar form to the ones in  case of quasi-elastic scattering, but there is an additional factor related to  the resonance excitation.  Each of the form factors has one free parameter. As a default for Nuance used in MiniBooNE  we use following values: $M_A=1.1~GeV$ for the axial-vector form factor and $M_V=0.84~GeV$ for the vector form factor.

The discrepancy in $Q^2$  for the \ccpip sample  is a long standing problem and the main goal of this paper is to understand it better. In the first step we will modify the Rein-Sehgal model \cite{RS81} of resonant pion production to include muon mass effects and new form factors. With this new model of pion production we will  present predictions for $CC1\pi^+$ production in the MiniBooNE detector with various parametrizations of the axial vector form factor. For the coherent process  the effect of mass of final lepton is included by applying the Adler factor  \cite{Adler}.

In the MiniBooNE detector  the struck nucleon is generally below the Cherenkov threshold. That is why the \ccpip sample consists of events  produced in following reactions 

\begin{eqnarray}
\nu + p &\to& \mu^-  +  p + \pi^+ (resonant\ \ production) \\
\nu + n &\to& \mu^-  + n  + \pi^+ (resonant\ \ production) \\
\nu +A &\to& \mu^- +A + \pi^+ \ \ (coherent\ \ production)
\end{eqnarray}

According to the Monte Carlo prediction for the MiniBooNE neutrino beam the biggest contribution comes from the excitation of the $\Delta(1234)$ resonance but the noresonant background and coherent production together account for about 10\% of the sample. In  the \ccpip sample 23\% of interactions occur on free protons, 46\% on bound  protons (eq. 1) and 12\% on bound neutrons (eq. 2). Remanning events are coherent production (6\%, eq.3) and various background interactions (13\%).

In the $CC1\pi^+$ reconstruction  only one track is reconstructed and assumed to be the $\mu^-$. The muon vertex, track angle, and energy, are found with a maximal likelihood fit, with the energy being determined from the total tank charge \cite{Patterson:2009ki}. The neutrino energy and four-momentum transfer are reconstructed from the observed muon kinematics, treating the interaction as a 2-body collision and assuming that the target nucleon is at rest inside the nucleus and that final hadronic state is of the $\Delta(1232)$ resonance.

\section{Final charged lepton  mass}

In the Rein-Sehgal  models of resonant and coherent  pion production via charged current the final  lepton is massless. Recently, three groups presented extensions of the RS model  of resonant production  with final charged lepton mass included.  Kuzmin, Lyubushlin and Naumov \cite{kln} (KLN model)  showed how one can include lepton mass effects from the lepton current.  In the Berger and Sehgal  model (BRS) \cite{brs}  in addition to the modification from the  KLN model the pion-pole contribution is added.  We modified our Monte Carlo generator to include the combined  KLN and BRS models. Predictions for KLN and BRS models are consistent with an  independent calculation with the lepton current modification and pion-pole contribution by Graczyk and Sobczyk \cite{Graczyk:2007xk}.

For models which  introduce a final lepton mass partial cross sections depend on the helicity $\lambda$
\begin{equation}
\frac{d \sigma}{dQ^2 dW^2} = \frac{G_F^2 \cos \theta_C}{8\pi^2M_N}\kappa \frac{Q^2}{|\bf q|^2} 
\sum \limits_{\lambda= \pm}\left [ \left( c_L^{\lambda}\right)^2 \sigma_L^{(\lambda)} 
+\left( c_R^{\lambda}\right)^2 \sigma_R^{(\lambda)}  + \left( c_S^{\lambda}\right)^2 \sigma_S^{(\lambda)} \right  ]
\end{equation}
KNL showed the form of  $c_{L,R,S}^{\lambda}$ for lepton current with the lepton mass.  In the limit of massless lepton in the final state,  ($m_l\to 0$), the cross section is reduced to the one in FRK model  ($c_L^{(-)}\to u$, $c_R^{(-)}\to v$, $c_S^{(-)}\to 2uv$ and $c_{L,R,S}^{(+)}\to 0$, eq. 2.18 in \cite{ravndal73}) . The helicity amplitudes $\sigma_{L,R,S}^{(\lambda)}$ are  calculated  as in  the FKR model but with three dynamical form factors (S, B and C) are modified.

An axial hadronic current has, in addition to the quark current $A_\mu$,  a pion-pole contribution from the PCAC hypothesis, which modifies the axial current.  In the case of a massless lepton in the final state the additional  term vanishes when contracted with lepton current. 
 
The effect of the lepton  mass in the leptonic  current is reached by changing only three out of seven dynamical form factors (KNL model). Subsequent modification due to the pion-pole term requires   alteration  of two dynamical form factors already modified \cite{brs}.

 \begin{eqnarray}
S\to  S_{KLN} ^\ld &&  \\
B\to  B_{KLN} ^\ld & \to &B_{BRS} ^\ld  \\
C\to  C_{KLN}^\ld &\to & C_{BRS} ^\ld
\end{eqnarray}

The coherent production in Nuance is described by the Rein-Sehgal model based on  by the Adler PCAC theorem \cite{pcac} describing neutrino scattering in the forward direction.  The extrapolation of the PCAC formula to non-forward angles is given by a slowly varying form-factor  with $M_A\approx 1GeV$.  

The important  modification of the cross section is due to the final lepton mass. This modification can be found in the paper by Adler \cite{Adler} and can be expressed as a simple multiplicative correction factor
\begin{equation}
{\cal C} = \left( 1- \frac{1}{2} \frac{Q^2_{min}}{Q^2+m_\pi^2}\right)^2
+\frac{1}{4}y \frac{Q^2_{min}\left(Q^2-Q^2_{min} \right)} {(Q^2+m_\pi^2)^2}
\end{equation} 
where
\begin{equation}
Q^2_{min}=m_l^2\frac{y}{1-y}, \ \ \ \ \ \ \ \ \  Q^2_{min}\leq Q^2\leq 2MEy_{max}
\end{equation}
where $y_{min}=m_\pi/E$ and $y_{max}=1-m_l/E$.

\section{Form factors}

The cross section in the Rein-Sehgal model is expressed in terms of helicity amplitudes but recent analysis on the vector and axial forma factors have been  presented in terms of form factors in the Rarita-Schwinger model. The connection between those two  formalism has been provided by Graczyk and Sobczyk  \cite{GS_ff}, who showed that  the vector form factor $G_V$ in the RS model can be expressed  in term of  $C_3^V$, $C_4^V$ and $C_5^V$

\begin{equation}
G_V^{new}=\frac{1}{2}\left( 1+\frac{Q^2}{\left( M+W\right)^2}\right)^{{1\over 2}-N} \sqrt{3(G_3^V)^2+(G_1^V)^2} \label{eq::new_gv}
\end{equation}
where
\begin{eqnarray}
G_3^V &=& \frac{1}{2\sqrt 3}\left[ C_4V\frac{W^2-Q^2-M^2}{2M^2} + C_5^V\frac{W^2+Q^2-M^2}{2M^2}
+C_3^V \frac{W+M}{M} \right] \\
G_1^V &=& -\frac{1}{2\sqrt 3}\left[ C_4V\frac{W^2-Q^2-M^2}{2M^2} + C_5^V\frac{W^2+Q^2-M^2}{2M^2}
-C_3^V \frac{M^2+Q^2+MW}{MW} \right] 
\end{eqnarray}

For our analysis we adopted vector form factors obtained by  Lalakulich et al. \cite{Lalakulich:2006sw}
\begin{eqnarray}
C_3^V &=&  2.13\left( 1 + \frac{Q^2}{4M_V^2} \right)^{-1}\left( 1 + \frac{Q^2}{M_V^2} \right)^{-2}\\
C_4^V &=& -1.51\left( 1 + \frac{Q^2}{4M_V^2} \right)^{-1}\left( 1 + \frac{Q^2}{M_V^2} \right)^{-2}\\
C_5^V &=&  0.48\left( 1 + \frac{Q^2}{4M_V^2} \right)^{-1}\left( 1 + \frac{Q^2}{0.766 M_V^2} \right)^{-2}
\end{eqnarray}

The limit  $Q^2 \to 0$ gives  $G_V^{new}(0)= 1.285$. This is higher than the value  assumed  in the RS model.  
The difference comes from the fact that with above shown choice of form factors $C_V^i$ it is not possible to reproduce the quark model prediction of vanishing electric contribution \cite{GS_ff}.

In order to calculate  similar  relations for the axial vector form factors it was necessary to assume the Adler model \cite{Adler:1968tw}
\begin{eqnarray} 
C_4^A &=& -\frac{1}{4} C_5^A\\
C_3^A &=& 0
\end{eqnarray}

Then $G_A$ and $C_5^A$ are related  by \cite{GS_ff}
\begin{equation}
\bar G_A^{new} =\frac{\sqrt 3}{2}\left( 1+\frac{Q^2}{\left( M+W\right)^2}\right)^{{1\over 2}-N}
\left[ 1- \frac{W^2-Q^2-M^2}{8M^2}\right]C_5^A \label{eq:c5gs}
\end{equation}

where the coupling constant $C_5^A(0)$ is predicted by the PCAC  
\begin{equation}
C_5^A(0) = \frac{g_{\Delta N \pi} f_\pi}{\sqrt 6 M} \approx 1.2.
\end{equation}

The values of this coupling constant  obtained in various quark models are lower than those determined from data.  The coupling constant evaluation is described in \cite{AlvarezRuso:1998hi}

Using the formula (\ref{eq:c5gs})  we checked these  parametrizations in the Rein-Sehgal model in the Nuance generator.

In the limit $Q^2 \to 0$ eq. (\ref{eq:c5gs}) is reduced to the  following relation
\begin{equation}
G_A(0)=\frac{\sqrt{3}} {2}\left (1-\frac{W^2-M^2}{8M^2} \right)C_5^A(0),
\end{equation}
which with $W=M_\Delta$ gives  $C_5^A(0)= 0.97$.

\begin{figure}[h]
\begin{tabular}{cc}
   \includegraphics[scale=.7]{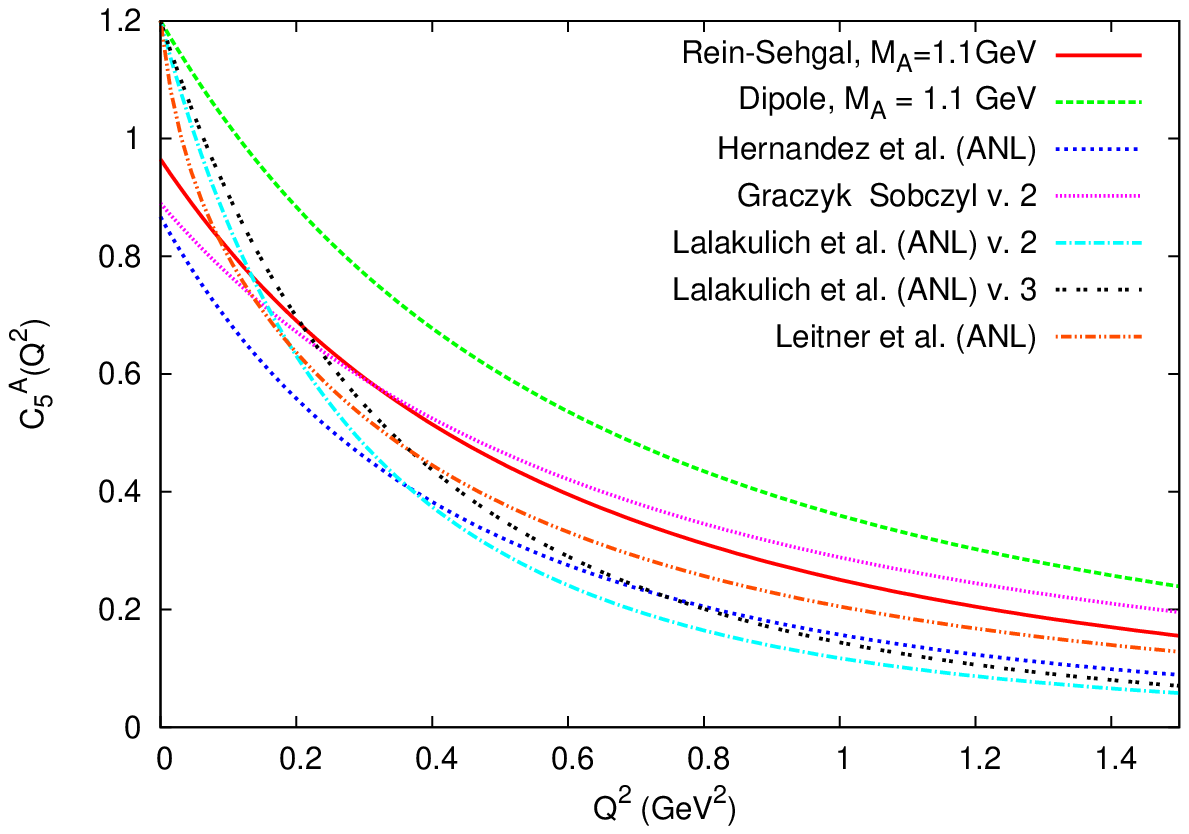}
&\includegraphics[scale=.7]{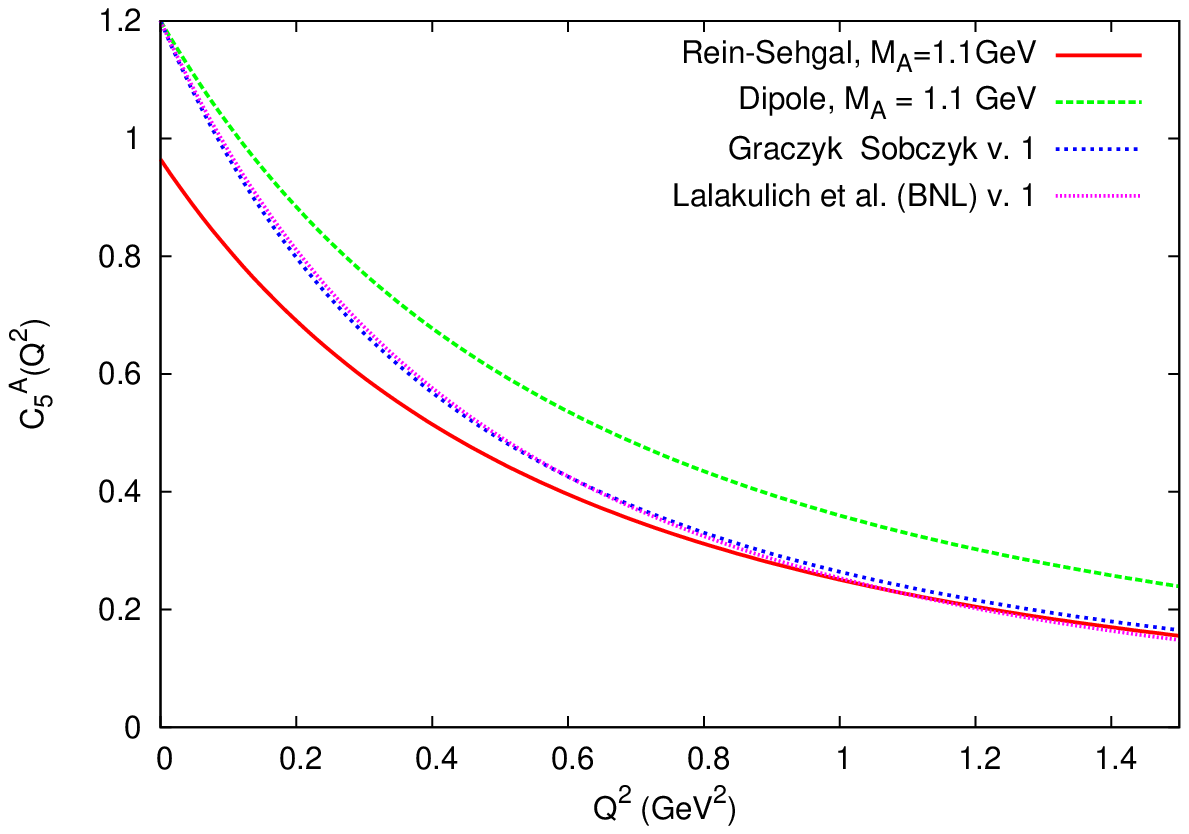}
\end{tabular}
\caption{The axial vector form factor $C_5^A$ for different parametrizations.  \label{fig::ga_var}. For the form factor in the Rein-Seghal model we used the eq. \ref{eq:c5gs} with $W=M_\Delta$.  Left: the parametrizaions which agree better with the ANL data compared to the RS form factor. Right: form factors which agree better with the BNL data with the RS form factor and dipole $C_5^A$ with $M_A=1.1 GeV$.    Notice that $C_5^A$ for the Lalakulich and Graczyk and Sobczyk v1 overlap.   }
\end{figure}

The $C_5^A(Q^2)$ form factor differs in the values of  $C_5^A(0)$, axial mass $M_A$ and the parametrization. Here we present predictions using  results obtained by three groups which  fit the existing data to the axial vector form factor in the Rarita-Schwinger formalism. 

\begin{description}

\item[Lalakulich et al. :] In the results by Lalakulich et al.  \cite{Lalakulich:2005cs} the axial form factor has been determined using the model of $\Delta^{++}(1232)$ resonance production with contributions only  from the  amplitude with isospin $3/2$. \\

In the paper three parametrizations have been presented. For the BNL data the axial form factor from  eq. \ref{lal1} has been used (ver. 1) with good agreement with the measured distribution for $Q^2>0.2GeV^2$. For the ANL data this from factor  overestimates  pion production so two other axial form factors were assumed (eqs. \ref{lal2} and \ref{lal3}) and the axial mass refitted.   

\begin{eqnarray}
ver.1: \ \ \ \ C_5^A = C_5^A(0) \left(1+\frac{Q^2}{M_A^2} \right)^{-2}\left(1+   \frac{Q^2}{3M_A^2} \right)^{-1}, & & C_5^A(0)=1.2,\ \ \ M_A=1.05 GeV \label{lal1} \\
ver.2: \ \ \ \ C_5^A = C_5^A(0) \left(1+\frac{Q^2}{M_A^2} \right)^{-2} \left(1+2\frac{Q^2}{  M_A^2} \right)^{-1}, & & C_5^A(0)=1.2,\ \ \ M_A=1.05 GeV  \label{lal2} \\
ver.3: \ \ \ \ C_5^A = C_5^A(0) \left(1+\frac{Q^2}{M_A^2} \right)^{-2}\left(1+   \frac{Q^2}{3M_A^2} \right)^{-2}, &  & C_5^A(0)=1.2,\ \ \ M_A=0.95 GeV \label{lal3}
\end{eqnarray}

\item[Hernandez et al.:] In the model described in \cite{Hernandez:2007qq}  in addition to the pion production due to $\Delta(1232)$ resonance excitation the background terms required by chiral symmetry are included. Assuming the parametrization in eq. \ref{her}   refitting  to the data was necessary since additional background terms changed the normalization. A  lower value of the axial mass $M_A$ was determined and the correction of the off-diagonal Goldberg-Treiman relation  was also lower of  about 30\% (eq. \ref{her}).

\begin{equation}
C_5^A = C_5^A(0)\left(1+\frac{Q^2}{M_A^2} \right)^{-2}\left(1+\frac{Q^2}{3 M_A^2} \right)^{-1}, \ \ \  C_5^A(0)=0.867,\ \ \ M_A=0.985 GeV \label{her}
\end{equation}

 \item[Graczyk and Sobczyk: ] The Rein-Sehgal model has been modified by the authors in order to determine  the $C_5^A$ form factor \cite{GS_ff} .  The fit  the dipole form of $C_5^A$  in the modified Rein-Sehgal model  using only the data for scattering on proton to avoid the background contribution.  They performed two fits for the both sets of data   simultaneously (ANL and BNL), one for axial mass only (eq. \ref{gs1}) and one for the axial mass and the $C_5^A(0)$ coupling constant (eq. \ref{gs2}). 
 
  \begin{eqnarray}
ver.1: \ \ \ \ C_5^A = C_5^A(0) \left(1+\frac{Q^2}{M_A^2} \right)^{-2}, & &C_5^A(0)=1.2, \  \ \  M_A=0.95~GeV \label{gs1} \\
ver.2: \ \ \ \ C_5^A = C_5^A(0) \left(1+\frac{Q^2}{M_A^2} \right)^{-2}, & &  C_5^A(0)=0.89,\ \ \  M_A= 1.15~ GeV^2 \label{gs2} 
\end{eqnarray}
 \end{description}

\begin{figure}[h]
   \includegraphics[scale=.7]{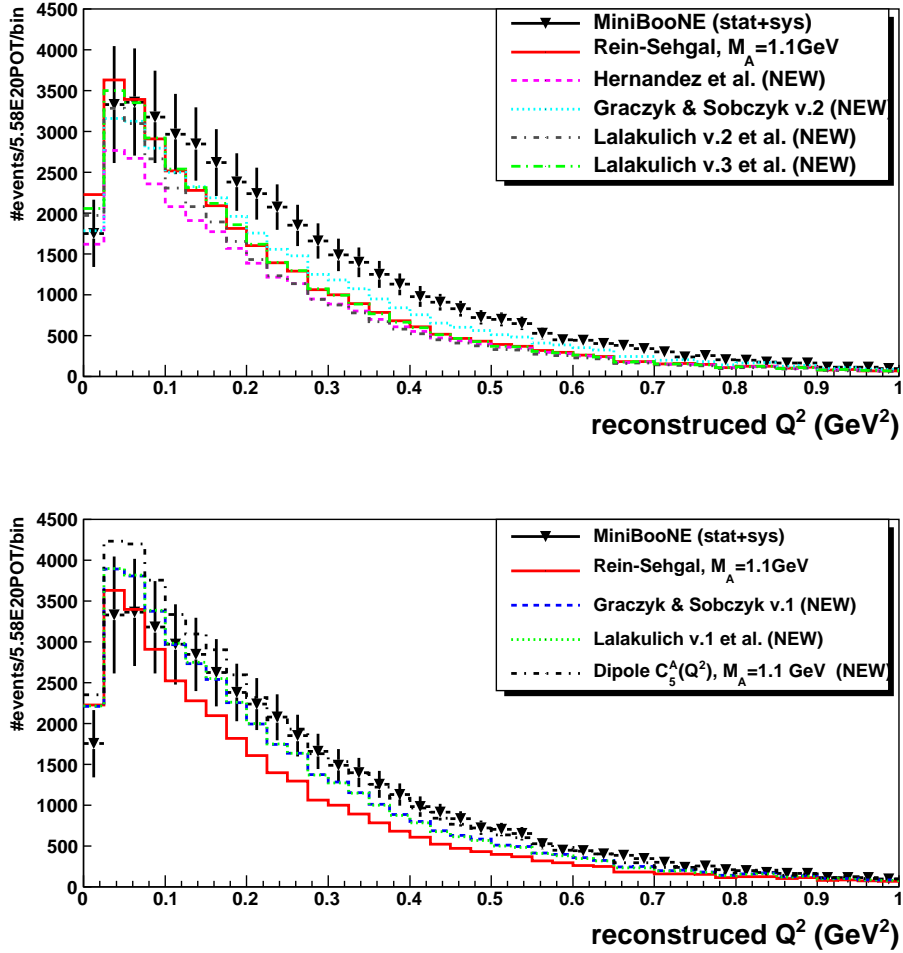}
\caption{MiniBooNE observed \ccpip events as a function of the four-momentum transfer compared with predictions. Top figure shows the  results with better agreement with the ANL data and bottom  figure with BNL data and dipole form of $C_5^A$ with $M_A=1.1GeV$.   For the bottom plot notice that the parametrization by Lalakulich et al. and  Graczyk  and Sobczyk are virtually the same and that the prediction for the  dipole form has good agreement with data. 
 \label{fig::q2_multi}}
\end{figure}

\section{Results}

In figure \ref{fig::ga_var} we present the  $Q^2$ dependency in the MiniBooNE data for the axial vector form factor $C_5^A$ parametrizations described above and the parametrization from Leitner et al. \cite{Leitner:2008ue}.  For better visualization we divided them into two groups, one with better agreement  with the ANL data and one with  better agreement with the BNL data. Each group is compared  with the RS form factor  obtained from the dipole form using eq. 18, and with dipole $C_5^A$ with $M_A=1.1 GeV$.   Notice that $C_5^A$ for the Lalakulich et al.  ver. 1 and Graczyk and Sobczyk ver. 1 overlap. The Rein-Sehgal parametrization was obtained by  applying eq. (\ref{eq:c5gs})  to the  dipole form of $C_5^A$ with $M_A=1.1~GeV$.

We have modified the original Rein-Sehgal model  in the Nuance generator  to include the mass effects and the new form factors discussed above (NEW).  The resulting predicted $Q^2$ distribution are compared to the MiniBooNE data in  figure \ref{fig::q2_multi}. We divided prediction into similar groups as in figure \ref{fig::ga_var}.  In the case of the parametrizations which agee with data from ANL  we can see good agreement for low $Q^2<0.1~ GeV^2$ and underproduction for rest of the region.  In case of BNL two parametrizations (Lalakulich ver.1 and Graczyk and Sobczyk ver. 1) which are almost identical but underproduction is not as big as for the ANL results. However, for the dipole form of the $C_5^A$ with $M_A=1.1~GeV$ agreement  for $Q^2>0.2GeV^2$ is very good  and there is deficit of observed \ccpip events for lower $Q^2$.

\end{document}